\documentstyle[array,multicol,eqsecnum,prd,aps,graphicx]{revtex}

\begin{document}
\draft

\preprint{STUPP-02-171}


\title{An SU(5) SUSY Model with $R$-Parity Violation\\
and Radiatively Induced Neutrino Masses }

\author{\bf Yoshio Koide\thanks{
On leave at CERN, Geneva, Switzerland.}
\thanks{
E-mail address: yoshio.koide@cern.ch; koide@u-shizuoka-ken.ac.jp}
and Joe Sato\thanks{E-mail address: joe@phy.saitama-u.ac.jp
}$^{(a)}$}
\address{
Department of Physics, University of Shizuoka, 
52-1 Yada, Shizuoka 422-8526, Japan \\
(a) Department of Physics, Faculty of Science, Saitama University,  
Saitama, 338-8570, Japan}

\date{\today}

\maketitle
\begin{abstract}
The radiatively induced neutrino mass matrix is investigated
on the basis of an SU(5) SUSY model.
In order  to evade the proton decay, an ansatz based on 
a discrete symmetry $Z_2$ is assumed: 
although, at the unification scale, we have two types of superfields 
$\Psi_{L(\pm)}=\overline{5}_{L(\pm)}+10_{L(\pm)}$, which  are 
transformed as
$\Psi_{L(\pm)} \rightarrow \pm \Psi_{L(\pm)}$ under the discrete 
symmetry Z$_2$, the particles $\Psi_{L(+)}$
are decoupled after the SU(5) symmetry is broken, so that 
our quarks and leptons belong to $\Psi_{L(-)}$.
The $R$-parity-violating terms for our quarks and leptons $\Psi_{L(-)}$
are basically forbidden under the symmetry Z$_2$.
However, we assume that mixings between members of $\Psi_{L(+)}$
and those of $\Psi_{L(-)}$ are in part caused after SU(5) is broken. 
As a result, the $R$-parity-violating interactions are in part
allowed, so that the neutrino masses are radiatively generated,
while the proton decay due to the $R$-parity violating terms is 
still forbidden because the term $d_R^c d_R^c u_R^c$ has $z=-1$.  
\end{abstract}


\maketitle

\begin{multicols}{2}

\narrowtext

\section{Introduction}

The origin of the neutrino mass generation is still a mysterious
problem in the unified understanding of the quarks and leptons.
The Zee model \cite{Zee} is one of several 
promising models, because it has only 3 free 
parameters and it can naturally lead to a large neutrino 
mixing \cite{Zee2}, especially, to a
bimaximal mixing \cite{Jarlskog}.
However,  the original Zee model is not on the framework of 
a grand unification theory (GUT). 
The most attractive idea  \cite{R_SUSY} to embed the Zee model 
into GUTs is to identify the Zee scalar  $h^+$ as the slepton 
$ \widetilde{e}_R$ in an $R$-parity-violating supersymmetric 
(SUSY) model. 
However, usually, it is accepted that SUSY models with $R$-parity 
violation are incompatible with a GUT scenario, 
because the $R$-parity-violating interactions induce 
proton decay \cite{Smirnov}.

In the present paper, in order to suppress this kind of proton 
decay, a discrete symmetry Z$_2$ 
is introduced.
The essential idea is as follows:
At the unification scale $\mu=M_X$, we have two types of superfields 
$\Psi_{L(\pm)}=\overline{5}_{L(\pm)}+10_{L(\pm)}$, which are 
transformed with 
$\pm 1$ under the discrete symmetry Z$_2$ (we will call it 
``Z$_2$-parity" hereafter).
We consider that the particles $\Psi_{L(+)}$
are decoupled after the SU(5) symmetry is broken (but Z$_2$ is
still unbroken), so that 
our quarks and leptons $\overline{5}_{L}+10_{L}$ belong to 
$\Psi_{L(-)}$.
The $R$-parity violating terms are given by the combinations
$\overline{5}_{(+)}\overline{5}_{(+)} 10_{(+)}$,
$\overline{5}_{(-)}\overline{5}_{(-)} 10_{(+)}$ and
$\overline{5}_{(+)}\overline{5}_{(-)} 10_{(-)}$,
so that they basically do not contribute to the quarks and
leptons with Z$_2$-parity $z=-1$, because of the Z$_2$ symmetry.
However, we assume that mixings of the members of $\Psi_{L(+)}$
with those of $\Psi_{L(-)}$ are caused in part after SU(5) is broken. 
As a result, the $R$-parity-violating interactions are in part
allowed, so that the neutrino masses are radiatively generated,
while the proton decay due to the $R$-parity-violating terms is 
still forbidden 
because the term $d_R^c d_R^c u_R^c$ is still exactly forbidden 
below $\mu=M_X$ in the present scheme.
The details will be discussed in the next section.

The purpose of the present paper is to investigate
the possible forms of the radiatively induced
neutrino mass matrix under the Z$_2$ symmetry.
In Sec.~III, we will give them, including a numerical study.
In Sec.~IV, we will give a comment on the Higgs scalars in the
present scheme.
Finally, Sec.~V is devoted to our conclusion.

\section{Z$_2$ symmetry and the proton decay} 
\label{sec:2}

We identify the Zee scalar $h^+$ as the slepton 
$\widetilde{e}_{R}^+$,
which is a member of SU(5) 10-plet sfermions 
$\widetilde{\psi}_{10}$. 
Then, the Zee interactions correspond to the following 
$R$-parity-violating interactions 
\begin{eqnarray}
\lambda_{ijk}(\overline{\psi}_{\overline{5}}^c)_i^A
(\psi_{\overline{5}})_j^B
(\widetilde{\psi}_{10})_{kAB}  \nonumber \\
=\frac{1}{\sqrt{2}}\lambda_{ijk}\left\{
\varepsilon_{\alpha \beta \gamma}
(\overline{d}_R)_i^{\alpha}(d_R^c)_j^{\beta}
(\widetilde{u}_R^{\dagger})_k^{\gamma} \right. \nonumber \\
-[(\overline{e}_L^c)_i(\nu_L)_j -
(\overline{\nu}_L^c)_i(e_L)_j]
(\widetilde{e}_R^{\dagger})_k \nonumber \\
- [(\overline{e}_L^c)_i(d_R^c)_j^{\alpha} -
(\overline{d}_R)_i^{\alpha}
(e_L)_j] (\widetilde{u}_L)_{k\alpha} \nonumber \\
\left.
+ [(\overline{\nu}_L^c)_i(d_R^c)_j^{\alpha} -
(\overline{d}_R)_i^{\alpha}
(\nu_L)_j](\widetilde{d}_L)_{k\alpha}\right\} \ , 
\end{eqnarray}
where $\psi^c\equiv C \overline{\psi}^T$ and the indices
$(i,j,\cdots)$, $(A,B,\cdots)$ and $(\alpha,\beta,\cdots)$
are family-, SU(5)$_{GUT}$- and SU(3)$_{colour}$-indices,
respectively.
The coefficients $\lambda_{ijk}$ are antisymmetric in
$i$ and $j$.
On the other hand, in SUSY GUT models, if the interactions (2.1) exist, 
the following $R$-parity-violating interactions will also 
exist:
\begin{eqnarray}
\lambda_{ijk}(\overline{\psi}_{\overline{5}}^c)_i^A
(\psi_{10})_{kAB}
(\widetilde{\psi}_{\overline{5}})_{j}^B \ ,\nonumber \\
=\frac{1}{\sqrt{2}}\lambda_{ijk}
\left\{ \varepsilon_{\alpha \beta \gamma}
(\overline{d}_R)_i^{\alpha}
(\widetilde{d}_R^{\dagger})_{j}^{\beta}
(u_R^{c})_k^{\gamma} \right.\nonumber \\
-[(\overline{e}_L^c)_i(\widetilde{\nu}_L)_j 
- (\overline{\nu}_L^c)_i
(\widetilde{e}_L)_j](e_R^c)_k \nonumber \\
- [(\overline{e}_L^c)_i(\widetilde{d}_
R^{\dagger})_j^{\alpha}
- (\overline{d}_R)_i^\alpha
(\widetilde{e}_L)_j] (u_L)_{k\alpha} \nonumber \\
\left.
+ [(\overline{\nu}_L^c)_i
(\widetilde{d}_R^{\dagger})_j^{\alpha} -
(\overline{d}_R)_i^{\alpha}
(\widetilde{\nu}_L)_j](d_L)_{k\alpha}
\right\} \ ,
\end{eqnarray}
which contribute to the proton
decay through the intermediate state
$\widetilde{d}_R$. 
Also, the term $(\overline{d}_R)_i^{\alpha}(d_R^c)_j^{\beta}
(\widetilde{u}_R^{\dagger})_k^{\gamma}$ in the interactions
(2.1) can contribute to the nucleon decay
through the intermediate state $\widetilde{u}_R$. 
The upper limits of the coupling constants 
$\lambda_{ijk}$ from proton decay experiments
have been investigated by Smirnov and Vissani \cite{Smirnov},
and the values must be highly suppressed.

In order to forbid the contribution
of the interactions (2.1) and (2.2)
to the proton decay, we  must consider that 
in the $R$-parity-violating interactions 
$\overline{5}\times \overline{5} \times 10$,
the term $ d_R^c d_R^c u_R^c$ is exactly forbidden,
while the terms $\nu_L e_L e_R^c$ and/or
$\nu_L d_R^c d_L$ are in part allowed.

For such purpose, we introduce a 
discrete symmetry Z$_2$, 
which exactly holds at every energy scale.
At the unification scale $\mu=M_X$, we have two types of superfields 
$\Psi_{L(\pm)}=\overline{5}_{L(\pm)}+10_{L(\pm)}$, which are 
transformed as $\Psi_{L(\pm)} \rightarrow \pm \Psi_{L(\pm)}$ 
under the discrete symmetry Z$_2$.
We consider that the particles $\Psi_{L(+)}$
are basically decoupled after the SU(5) symmetry is broken,
so that our quarks and leptons (and their SUSY partners) 
$\overline{5}_{L}+10_{L}$ are
regarded as $\Psi_{L(-)}=\overline{5}_{L(-)}+10_{L(-)}$.
The $R$-parity-violating terms for quarks and leptons (and their
SUSY partners) are basically forbidden under the
symmetry Z$_2$ below $\mu=M_X$, because the terms are composed of
$\overline{5}_{L(-)} \overline{5}_{L(-)} 10_{L(-)}$.

However, if we assume that  mixings between the members of $\Psi_{L(+)}$
and those of $\Psi_{L(-)}$ in part take place after SU(5) is broken,
$R$-parity-violating interactions $\Psi_{L(+)} \Psi_{L(-)} \Psi_{(-)}$
become available at the low energy $\mu=m_Z$, too. 
For example, we assume  a mixing
\begin{equation}
(2,1)_{Li} = (2,1)_{L(-)i} \cos\theta_i^A +(2,1)_{L(+)i}
\sin\theta_i^A \ ,
\end{equation}
between the $(2,1)$ components of SU(2)$\times$SU(3)  
for the $i$-th family.
(Hereafter, we will refer to the mixing (2.3) as a mixing of type 
A$_i$.)
Then, the $R$-parity-violating interactions 
\begin{eqnarray}
& & 
\sin\theta_i^A \lambda_{ijk} \nu_{Li} d_{Rj}^c d_{Lk} \ , \ \ 
\sin\theta_i^A \lambda_{ijk} e_{Li} d_{Rj}^c u_{Lk} \ ,
\nonumber \\
& &
\sin\theta_i^A \cos\theta_j^A \lambda_{ijk} \nu_{Li} e_{Lj} e_{Rk}^c \ ,
\end{eqnarray}
become available from the interactions
\begin{equation} 
\lambda_{ijk} \overline{5}_{(+)i} \overline{5}_{(-)j} 10_{(-)k} \ ,
\end{equation}
above the unification scale $\mu=M_X$.
Also, we can consider a mixing
\begin{equation}
(2,3)_{Lk} = (2,3)_{L(-)k} \cos\theta_k^B +(2,3)_{L(+)k}
\sin\theta_k^B \ ,
\end{equation}
between the $(2,3)$ components of SU(2)$\times$SU(3)  
for the $k$-th family.
(Hereafter, we will refer to the mixing (2.6) as a B$_k$-type
mixing.)
Then, the $R$-parity-violating interactions
\begin{equation}
\sin\theta_k^B \lambda'_{ijk} \nu_{Li} d_{Rj}^c d_{Lk}  \ {\rm and}  \ 
\sin\theta_k^B \lambda'_{ijk} e_{Li} d_{Rj}^c u_{Lk}  
\end{equation}
become available from the interactions 
\begin{equation} 
\lambda'_{ijk}  \overline{5}_{(-)i} \overline{5}_{(-)j} 
10_{(+)k} \ .
\end{equation}
On the other hand, note that the interaction
\begin{equation}
d_R^c d_R^c u_R^c  
\end{equation}
is exactly forbidden, independently of whether the mixings (2.3)
and (2.6)  occur or not, because those interactions have
the Z$_2$ parity $z=-1$.
Therefore, the proton decay due to the $R$-parity-violating 
terms is exactly forbidden 
because of the absence of the term $d_R^c d_R^c u_R^c$.
On the other hand, 
the neutrino masses are radiatively generated
through the interactions $\nu_L d_R^c d_L$ and
$\nu_L e_L e_R^c$  with $z=+1$.
The possible forms of the radiative neutrino mass
matrix will be discussed in the next section.

At present, we do not know a reasonable mechanism not only for 
such a mixing, but also for the decoupling of $\Psi_{L(+)}$.
In order to make $\Psi_{L(+)}=\overline{5}_{L(+)}+10_{L(+)}$ heavy,
the SU(2)$_L$ symmetry must be broken, but, of course, we cannot 
consider a scenario in which SU(2) is broken just after SU(5) is 
broken.
In the present paper, we give only a phenomenological selection rule: 
if the superfield $\Psi$ can make a five-body SU(5) singlet operator 
$\Psi \Psi \Psi \Psi \Psi$ with the Z$_2$ parity $z=+1$, 
then the superfield $\Psi$ can be decoupled below $\mu=M_X$. 
Obviously, according to this selection rule,
the superfield $\overline{5}_{L(+)}$ can be decoupled below
$\mu=M_X$.
Similarly, the superfield $10_{L(+)}$ is decoupled below $\mu=M_X$.
However, note that those operators in the SU(5) singlets  are 
symbolically expressed in terms of SU(2)$\times$SU(3) components 
as follows:
\begin{eqnarray}
\left(\overline{5}_{L(+)}\right)^5
& =& [(2,1)_{L(+)}]^2\times [(1,3)_{L(+)}]^3 \ , \\
\left(10_{L(+)}\right)^5
& =& (1,1)_{L(+)}\times[(1,\overline{3})_{L(+)}]^2\times 
[(2,3)_{L(+)}]^2 \nonumber \\
& & +(1,\overline{3})_{L(+)} \times [(2,3)_{L(+)}]^4 \ ,
\end{eqnarray}
and that even if the interchanges 
$(2,1)_{L(+)i}  \leftrightarrow  (2,1)_{L(-)i}$, 
and/or 
$(2,3)_{L(+)k}  \leftrightarrow  (2,3)_{L(-)k}$,
are caused, the composite operators 
\begin{equation}
\left(\overline{5}_{L}\right)^5
 =  [(2,1)_{L(-)}]^2\times [(1,3)_{L(+)}]^3 \ , 
\end{equation}
\begin{eqnarray}
\left(10_{L}\right)^5
& =& (1,1)_{L(+)}\times[(1,\overline{3})_{L(+)}]^2\times 
[(2,3)_{L(-)}]^2  \nonumber  \\
& & +(1,\overline{3})_{L(+)} \times [(2,3)_{L(-)}]^4 
\end{eqnarray}
still have $z=+1$.
Such interchanges are possible only for the components
$(2,1)_L$ and $(2,3)_L$.
As a result, only  the combination
\begin{eqnarray}
\overline{5}_{L} +10_L = [(2,1)_{L(+)} + (1,\overline{3})_{L(-)}]
\nonumber \\
+[(1,1)_{L(-)}+(2,3)_{L(-)}+(1,\overline{3})_{L(-)}] \ ,
\end{eqnarray}
for the $i$-th family and/or
\begin{eqnarray}
\overline{5}_L +10_L =
[(2,1)_{L(-)} + (1,\overline{3})_{L(-)}] 
\nonumber \\
+[(1,1)_{L(-)}+(2,3)_{L(+)}+(1,\overline{3})_{L(-)}] \ , 
\end{eqnarray}
for the $k$-th family
survive below $\mu=M_X$ as the quarks and leptons
(and their SUSY partners).

Of course, the above selection rule cannot be justified within
the framework of the minimal SUSY standard model.
At present, this is only an ansatz to select which components of
SU(2)$\times$SU(3) can be interchanged.


\section{Radiative neutrino masses}
\label{sec:3}

In a SUSY GUT scenario, there are many origins of the 
neutrino mass generations.
For example, the sneutrinos $\widetilde{\nu}_{iL}$ can
have vacuum expectation values (VEVs), and  
the neutrinos $\nu_{Li}$ acquire
their masses thereby (for example, see Ref.\cite{Diaz}).
Although we cannot rule out a possibility that the observed
neutrino masses can be understood from such compound origins,
we do not take such a point of view in the present paper, 
because the observed neutrino masses and mixings appear to be 
rather simple and characteristic.
We simply assume that 
the radiative masses are only dominated even if there are
other origins of the neutrino mass generations.

In the present scenario, the origins of the radiatively
induced neutrino masses are two: one is induced
by the $R$-parity-violating interactions
$\nu_L d_R^c \widetilde{d}_L$ and $\nu_L \widetilde{d}_R^c
d_L$; the other one is induced by $\nu_L e_L \widetilde{e}_R^c$ 
and $\nu_L \widetilde{e}_L e_R^c$. 
Note that there is no Zee-type diagrams due to 
$H_d^+$--$\widetilde{e}^+_{R}$ mixing in this scheme.

First, we discuss the down-quark loop contributions.
For simplicity, we assume that the masses $\widetilde{M}_{Li}$
and $\widetilde{M}_{Ri}$ of the squarks
$\widetilde{d}_{Li}$ and $\widetilde{d}_{Ri}$ are approximately 
constant, independently of the flavours, although we
consider the flavour-dependent structure for the mass terms
$\widetilde{d}_L^\dagger \widetilde{M}_d^2 \widetilde{d}_R$.
Then, the radiatively induced neutrino mass matrix due to
the A-type mixing is given by
\begin{equation}
(M_\nu)_{ij} = m_0 \lambda_{ikm} \lambda_{jln}
(M_d^\dagger)_{kn} (\widetilde{M}_d^{2\dagger})_{lm}
+(i\leftrightarrow j) \ ,
\end{equation}
where sine-factors have been drooped for simplicity, 
$M_\nu$ is defined by $\overline{\nu}_L M_\nu \nu_L^c$, 
and the coupling constants $\lambda_{ijk}$ 
are redefined by
\begin{equation}
\lambda_{ijk} \left[ \overline{\nu}_{Li} d_{Rj} 
\widetilde{d}_{Lk}^\dagger + \overline{\nu}_{Li}
\widetilde{d}_{Rj} d_{Lk}^c -(\nu_L \leftrightarrow
d_R^c) \right] \ .
\end{equation}
Here, we have changed the definition of $\lambda_{ijk}$
from that in (2.1) 
as $\lambda_{ijk} \rightarrow \lambda_{ijk}^*$
for the convenience of the expression of $M_\nu$
defined by $\overline{\nu}_L M_\nu \nu_L^c$.
In the present paper, the unitary matrix $U_\nu$ 
used to diagonalize the Majorana mass matrix $M_\nu$ is 
defined as $U_\nu^\dagger M_\nu U_\nu^* = D_\nu$.
Then,  the so-called Maki--Nakagawa--Sakata--Pontecorvo 
\cite{MNS}  matrix (we will simply call it the 
``lepton mixing matrix") $U \equiv U_{MNSP}$ is
given by $U= U_L^{e\dagger} U_\nu$.
Usually, it is considered that the matrix form of 
$\widetilde{M}_d^2$ is proportional to the form $M_d$. 
Then, the neutrino mass matrix (3.1) becomes, in a more
concise form:
\begin{equation}
(M_\nu)_{ij} = m_0 \lambda_{ikm} \lambda_{jln}
(M_d^\dagger)_{kn} (M_d^\dagger)_{lm} \ ,
\end{equation}
where we have redefined the common factor $m_0$ from
that in (3.1).
Of course, exactly speaking, for the A$_i$ mixings with 
$\sin\theta_i^A$-factors, 
we should read the expression (3.3) as 
\begin{equation}
(M_\nu)_{ij} = m_0 s_i^A s_j^A \lambda_{ikm} \lambda_{jln}
(M_d^\dagger)_{kn} (M_d^\dagger)_{lm} \ ,
\end{equation}
where $s_i^A=\sin\theta_i^A$ defined in (2.3).
When we consider the B$_k$ mixings, we read the expression
(3.3) as 
\begin{equation}
(M_\nu)_{ij} = m_0 s_m^B s_n^B \lambda'_{ikm} \lambda'_{jln}
(M_d^\dagger)_{kn} (M_d^\dagger)_{lm} \ ,
\end{equation}
where $s_k^B=\sin\theta_k^B$ defined in (2.6).
For mixed-type mixings of A$_i$ and B$_k$, we read (3.3)
as
\begin{eqnarray}
(M_\nu)_{ij}& = & m_0 
(s_i^A c_m^B \lambda_{ikm} + c_i^A s_m^B\lambda'_{ikm})\nonumber \\
 & & (s_j^A c_n^B \lambda_{jln}+ c_j^A s_n^B\lambda'_{jln})
 (M_d^\dagger)_{kn} (M_d^\dagger)_{lm} \ ,
\end{eqnarray}
where $c_i^A=\cos\theta_i^A$ and $c_k^B=\cos\theta_k^B$. 

The contributions from the charged lepton loops are essentially
the same as (3.3), except for the absence of the B-type mixing and
the replacement $M_d \rightarrow M_e^T$.
For simplicity, we will continue the investigation for the case of 
the down-quark loop contributions.

For the phenomenological study of the mass matrix (3.3),
it is convenient to take the basis on which the 
down-quark mass matrix $M_d$ is diagonal:
\begin{equation}
U_L^{d\dagger} M_d U_R^d =D_d \equiv {\rm diag}
(m_1^d, m_2^d, m_3^d) \ .
\end{equation}
We consider that, on the basis with $M_d =D_d$, 
the charged lepton mass matrix $M_e$
is also approximately diagonal, $U_L^{e \dagger} M_e
U_R \simeq D_e = {\rm diag}(m_1^e, m_2^e, m_3^e)$,
so that the unitary matrix $U_\nu$  approximately
gives the lepton mixing matrix $U= U_L^{e\dagger}U_\nu$.
Then, we can express (3.3) as
\begin{eqnarray}
(M_\nu)_{11}& =& (m_3^d)^2 (\lambda_{133})^2
+(m_2^d)^2 (\lambda_{122})^2
+2 m_3^d m_2^d \lambda_{123}\lambda_{132} \ ,
\nonumber \\
(M_\nu)_{22}& =& (m_3^d)^2 (\lambda_{233})^2
+(m_1^d)^2 (\lambda_{211})^2
+2 m_3^d m_1^d \lambda_{213}\lambda_{231} \ ,
\nonumber \\
(M_\nu)_{33}& =& (m_2^d)^2 (\lambda_{322})^2
+(m_1^d)^2 (\lambda_{311})^2
+2 m_2^d m_1^d \lambda_{312}\lambda_{321} \ ,
\nonumber \\
(M_\nu)_{12}& =& (m_3^d)^2 \lambda_{133}\lambda_{233}
+ m_3^d m_2^d \lambda_{123}\lambda_{232} 
\nonumber \\
& & +  m_3^d m_1^d \lambda_{131}\lambda_{213} 
+ m_2^d m_1^d \lambda_{121}\lambda_{212} \ ,
\nonumber \\
(M_\nu)_{13}& =& (m_2^d)^2 \lambda_{122}\lambda_{322}
+ m_3^d m_2^d \lambda_{132}\lambda_{323} 
\nonumber \\
& &+  m_3^d m_1^d \lambda_{131}\lambda_{313} 
+ m_2^d m_1^d \lambda_{121}\lambda_{312} \ ,
\nonumber \\
(M_\nu)_{23}& =& (m_1^d)^2 \lambda_{211}\lambda_{311}
+ m_3^d m_2^d \lambda_{232}\lambda_{323} 
\nonumber \\
& & +  m_3^d m_1^d \lambda_{231}\lambda_{313} 
+ m_2^d m_1^d \lambda_{212}\lambda_{321} \ ,
\nonumber \\
\end{eqnarray}
where, for simplicity, we have dropped the common factor $m_0$.
In order to give the best-fit values for the observed neutrino data
\cite{solar,atm}
\begin{equation}
R \equiv \frac{\Delta m_{21}^2}{
\Delta m_{32}^2} 
\simeq \frac{5.0 \times 10^{-5} {\rm eV^2}}
{2.5 \times 10^{-3} {\rm eV^2}}
= 2.0 \times 10 ^{-2} \ ,
\end{equation}
\begin{eqnarray}
& & \sin^2 2\theta_{solar} =\sin^2 2\theta_{12}
= 0.76 \ , \nonumber \\
& & [\tan^2 \theta_{solar}=0.34] \ ,
\end{eqnarray}
\begin{equation}
\sin^2 2\theta_{atm} =\sin^2 2\theta_{23}
= 1.0 \ ,
\end{equation}
we must seek a parameter set that gives
$(M_\nu)_{22}\simeq (M_\nu)_{33}$ and
$(M_\nu)_{12}\sim (M_\nu)_{13}$ for the
expression (3.8). 
When we consider the A$_i$-type mixings, we obtain \cite{YS}
\begin{equation}
M_\nu \propto \left(
\begin{array}{ccc}
\varepsilon^2 & \varepsilon & \varepsilon \\
\varepsilon & 1 & 1 \\
\varepsilon & 1 & 1 
\end{array} \right) \ ,
\end{equation}
for $\varepsilon=s_1^A/s_2^A=s_1^A/s_3^A$ and 
$\lambda_{2j3}/\lambda_{3j2} \simeq m_2^d/m_3^d$.
Generally, when we consider only A$_i$-type mixings,
the solutions are highly dependent on the fine-tuning
among the coefficients $\lambda_{ijk}$, and, besides, since
the mass matrix is too near to a rank-1 matrix, it
is difficult to give the solution a small but sizeable value
of $R$ (it leads to an extremely small value of $R$).
Even if we take the charged lepton loop contributions into
consideration, the situation is not improved unless
we assume the special parametrization for $\lambda_{ijk}$
($\lambda_{2j3}/\lambda_{3j2} \simeq m_2^d/m_3^d$ and so on).

Next, we consider the case of the B$_k$-type mixings.
When  we consider a B$_1$ mixing, we obtain a simple
mass matrix form
\begin{equation}
M_\nu = (s_1^B)^2 (m_1^d)^2 \left(
\begin{array}{ccc}
0 & 0 & 0 \\
0 & (\lambda'_{211})^2 & \lambda'_{211}\lambda'_{311} \\
0 & \lambda'_{211}\lambda'_{311} & (\lambda'_{311})^2 
\end{array} \right) \ ,
\end{equation}
because the case gives the relations $\lambda'_{ij2}s_2^B
=\lambda'_{ij3}s_3^B =0$.
It is natural to consider that $\lambda'_{211} \simeq \lambda'_{311}$
since those come from the same interactions (2.8) [not from
(2.5)].  Therefore, the mass matrix (3.13) can give a
maximal mixing between $\nu_\mu$ and $\nu_\tau$.
A small additional term will reasonably give a bimaximal
mixing.

For example, we consider a mixed-type, 
A$_3$ and B$_1$, mixing.
In this case, the $\lambda_{ijk}$ in (3.3) must be replaced
according to Table I.
It is natural to consider that $\lambda_{ijk} \simeq {\rm const} 
\equiv \lambda$ and $\lambda'_{ijk} \simeq {\rm const} \equiv \lambda'$.
Then, the mass matrix $M_\nu$ is reduced to the form
\begin{equation}
M_\nu \simeq (s_1^B)^2 (m_1^d) \left(
\begin{array}{ccc}
0 & 0 & \varepsilon (r_3+r_2) \\
0 & 1 & 1 +\varepsilon r_3 \\
\varepsilon (r_3+r_2) & 1 +\varepsilon r_3 & 
(1+\varepsilon r_2)^2 
\end{array} \right) \ ,
\end{equation}
where $\varepsilon= s_3^A \lambda/s_1^B \lambda' \ll 1$, 
$r_2=m_2^d/m_1^d$ and $r_3=m_3^d/m_1^d$, and we have
assumed $c_3^A \simeq 1$.
This mass matrix can give a reasonable value of $R$
together with a nearly bimaximal mixing.
For example, for the parameter value $\varepsilon=0.000374$,
we obtain the following numerical results:
\begin{equation}
m_1^\nu= 0.0822 m_0, \ \ m_2^\nu=-0.3276 m_0, \ \ 
m_3^\nu =2.2604 m_0,
\end{equation}
\begin{equation}
U = \left(
\begin{array}{ccc}
0.8726 & -0.4822 & 0.0776 \\
-0.3925 & -0.5978 & 0.6990 \\
0.2907 & 0.6404 & 0.7109 
\end{array} \right) \ ,
\end{equation}
i.e.
\begin{equation}
R=0.0201 \ ,
\end{equation}
\begin{equation}
\sin^2 2\theta_{12}\equiv 4U_{11}^2 U_{12}^2 = 0.708 \ ,
\end{equation}
\begin{equation}
\sin^2 2\theta_{23}\equiv 4U_{23}^2 U_{33}^2 = 0.988 \ .
\end{equation}
These values are in good agreement with the 
best fit values (3.9)--(3.11) for the observed neutrino data.
Although  it is difficult in the original Zee model to give 
a sizeable
deviation of $\sin^2 2\theta_{12}$ from 1 \cite{koide-zee} 
(it must be $\sin^2 2\theta_{12}=1.0$), 
the present model can give a reasonable
deviation from  $\sin^2 2\theta_{12}=1.0$.
The result
\begin{equation}
U_{13}^2 = 0.00602 
\end{equation}
is also consistent with the present experimental upper limit
\begin{equation}
|U_{13}|^2 < 0.03 \ ,
\end{equation}
from the CHOOZ collaboration \cite{CHOOZ}.

\section{Higgs sectors}

In the present model, the quark and charged lepton mass matrices
are generated by the VEVs of the Higgs scalars with $z=+1$.
Therefore, even if SU(2)$_L$ is broken later, the Z$_2$
symmetry still exactly holds.

Let us show the  mass matrices $M_f$ for the case of the 
mixed-type mixing A$_3$ and $B_1$ as an example of the explicit
forms of $M_f$:
\begin{equation}
M_e = \left(
\begin{array}{ccc}
c_1 & b_1 & a_1 \\
c_2 & b_2 & a_2 \\
c_1^A c_3 & c_1^A b_3 & c_1^A a_3 
\end{array} \right) \ ,
\end{equation}
\begin{equation}
M_d =
\left(
\begin{array}{ccc}
c_1^B c_1 & c_1^B c_2 & c_1^B c_3 \\
b_1 & b_2 & b_3 \\
a_1 & a_2 & a_3 
\end{array} \right) \ , 
\end{equation}
\begin{equation}
M_u =
\left(
\begin{array}{ccc}
c_1^B c'_1 & c_1^B c'_2 & c_1^B c'_3 \\
b'_1 & b'_2 & b'_3 \\
a'_1 & a'_2 & a'_3 
\end{array} \right) \ . 
\end{equation}
Here, $M_f$  have been defined by 
$\overline{f}_L M_f f_R$ ($f=u,d,e$).
Note that the mass matrix $M_d$ has a form different
from  $M_e^T$ because of the mixing factors.
Usually, if we consider one type of  Higgs scalar of SU(5) 5-plet
($\overline{5}$-plet), it is difficult to obtain realistic
mass matrices $M_f$ ($f=u,d,e$).
Therefore, the present model has a possibility to improve
this problem.
However, whether we can give reasonable mass matrix forms
of $M_f$ or not is a future task to us.

Now, we would like to give a comment on the Higgs sectors.
In Sec.~II, we have assumed that although the superfield 
$\overline{5}_{L(+)}$ is decoupled below $\mu=M_X$, 
the components $(2,1)_{(+)}$ of $\overline{5}_{L(+)}$
can contribute to low energy phenomena through the mixing
(2.3).
If we consider the Higgs fields $\overline{H}_{d(+)}$
and $\overline{H}_{d(-)}$ with SU(5) $\overline{5}$-plet, 
and if we assume a situation similar to the matter fields
$\overline{5}_L$, then we obtain
\begin{equation}
\overline{H}_d = (2,1)_{(+)} + (1,\overline{3})_{(-)} \ ,
\end{equation}
where we have assumed a perfect interchange between
$(2,1)_{(-)}$ and $(2,1)_{(+)}$, not a mixing.
Note that in this scheme, the Higgs scalar component
$(3,1)_{(-)}$ cannot couple to the fermions 
$\overline{d}_R u_R^c$ with $z=+1$ independently of
the mixings A$_i$ and $B_k$, so that the scalar
$(3,1)_{(-)}$ cannot contribute to the proton decay
and it need not be super-heavy.
(Although it couples to the fermions 
$(\overline{\nu}_L^c d_L -\overline{e}_L^c u_L)$,
these interactions cannot contribute to the proton
decay.)
We can consider a similar mechanism for the Higgs fields
$H_u$ with SU(5) 5-plet.
The interactions of $(1,3)_{(-)}$ with $\overline{u}_L^c d_L$
and $\overline{e}_R u_R^c$ are absent, so that the scalar
$(1,3)_{(-)}$ need not be super-heavy.
The so-called $\mu$-terms are composed of
$H_{u(+)} \overline{H}_{d(+)}$ and
$H_{u(-)} \overline{H}_{d(-)}$.

\section{Conclusion}

In conclusion, we have investigated  possible forms of
radiatively induced neutrino mass matrix under an
ansatz [(2.3) and (2.6)] within the framework of the SU(5)
SUSY model.
We have assumed two types of matter fields $\Psi_L =
\overline{5}_L +10_L$, $\Psi_{(\pm)}$, which are transformed 
as $\Psi_{(\pm)} \rightarrow \pm \Psi_{(\pm)}$
under the Z$_2$ symmetry.
We have assumed that the Z$_2$ symmetry exactly holds,
even if SU(5) is broken.
The essential ansatz is in the mixings (2.3) and (2.6).
Although the origin of the mixings (2.3) and (2.6) is
still an open question, if we admit this ansatz, we can obtain 
very interesting and simple neutrino mass matrix form, which 
can give satisfactory numerical results for the observed 
neutrino data.
How we can justify the ansatz is a future task for us.

\vspace*{3mm}
\begin{center}
{\large\bf Acknowledgements}
\end{center}

The work based on the Z$_2$ symmetry was first begun by
Dr.~A.~Ghosal and one of the authors (Y. K.), but the
earlier version \cite{Z2} (hep-ph/0203113) failed
to suppress the proton decay sufficiently.
This work is also based on the Z$_2$ symmetry, but
the suppression mechanism is completely different from
the previous one.  Y. K. would like to thank Dr.~A.~Ghosal
for his collaboration at the earlier stage.
Y. K. also wishes to acknowledge the
hospitality of the Theory group at CERN, where this
work was completed.
The work of J. S. is supported in part by Grants-in-Aid for 
Scientific Research from the Ministry of Education, Science, Sports, 
and Culture of Japan, No.14740168, No.14039209 and No.14046217.


\vspace{5mm}
\begin{table}
\begin{center}

\begin{tabular}{|l|c|c|c|}\hline
 $i$    & $k=1$        & $k=2$     & $k=3$ \\ \hline
$1$ & $s_1^B \lambda'_{1j1}$ & 0 & 0 \\
$2$ & $s_1^B \lambda'_{2j1}$ & 0 & 0 \\
$3$ & \begin{tabular}{c}
$c_3^A s_1^B \lambda'_{3j1}$ \\
+$s_3^A c_1^B \lambda_{3j1}$ 
\end{tabular} & $s_3^A \lambda_{3j2}$  &
$s_3^A  \lambda_{3j3}$ \\ \hline
\end{tabular}
\vspace{2mm}
\caption{ 
Rule  of the replacement $\lambda_{ijk}$ in the mass matrix (3.3) 
for the case of A$_3$ and B$_1$ mixings.
}
\end{center}

\label{A3B1}

\end{table}
\vspace{2mm}

\end{multicols}

\end{document}